\documentclass[aps,prl,twocolumn,preprintnumbers,amsmath,amssymb,superscriptaddress]{revtex4-2}
\usepackage{amsmath,graphicx}
\usepackage[utf8]{inputenc}
\usepackage[T1]{fontenc}
\usepackage{xcolor}
\usepackage{textcomp}
\usepackage{bm}

\usepackage{siunitx}
\usepackage{physics}
\usepackage{amsmath}
\usepackage{tikz}
\usepackage{mathdots}
\usepackage{yhmath}
\usepackage{cancel}
\usepackage{color}
\usepackage{array}
\usepackage{multirow}
\usepackage{amssymb}
\usepackage{gensymb}
\usepackage{tabularx}
\usepackage{extarrows}
\usepackage{booktabs}
\usetikzlibrary{fadings}
\usetikzlibrary{patterns}
\usetikzlibrary{shadows.blur}
\usetikzlibrary{shapes}

\usepackage{color}
\definecolor{LinkColor}{rgb}{0.75,0.0,0.2}

\usepackage{hyperref}
\hypersetup{
	pdfauthor={good guys},
	pdftitle={good title},
	colorlinks=true,
	citecolor=LinkColor,
	linkcolor=LinkColor,
	urlcolor=LinkColor,
}

\usepackage{listings}
\definecolor{lightgray}{gray}{1}

\usepackage{theorem}

\def\Eq#1{Eq.~(\ref{#1})}

\usepackage{tikz}

\newcommand{\nc}{\newcommand}
\nc{\braoprket}[3]{\langle#1|#2|#3\rangle}
\nc{\opn}[1]{\operatorname{#1}}
\nc{\avg}[1]{\langle#1\rangle}
\nc{\ketbrasame}[1]{|#1\rangle\!\langle#1|}
\nc{\swap}{\opn{SWAP}}
\nc{\E}{\mathbb{E}}
\nc{\Var}{\opn{Var}}
\nc{\dg}{\dagger}

\usepackage[normalem]{ulem}
\nc{\hknew}[1]{\textcolor{brown}{#1}}

\begin{document}
\title{Entanglement Growth from Entangled States: A Unified Perspective on Entanglement Generation and Transport}

\author{Chun-Yue Zhang}
\affiliation{Beijing National Laboratory for Condensed Matter Physics $\&$ Institute of Physics, Chinese Academy of Sciences, Beijing 100190, China}
\affiliation{University of Chinese Academy of Sciences, Beijing 100049, China}

\author{Zi-Xiang Li}
\email{zixiangli@iphy.ac.cn}
\affiliation{Beijing National Laboratory for Condensed Matter Physics $\&$ Institute of Physics, Chinese Academy of Sciences, Beijing 100190, China}
\affiliation{University of Chinese Academy of Sciences, Beijing 100049, China}

\author{Shi-Xin Zhang}
\email{shixinzhang@iphy.ac.cn}
\affiliation{Beijing National Laboratory for Condensed Matter Physics $\&$ Institute of Physics, Chinese Academy of Sciences, Beijing 100190, China}

\date{\today}

\begin{abstract}
Studies of entanglement dynamics in quantum many-body systems have focused largely on initial product states. Here, we investigate the far richer dynamics from initial entangled states, uncovering universal patterns across diverse systems ranging from many-body localization (MBL) to random quantum circuits. Our central finding is that the growth of entanglement entropy can exhibit a counter-intuitive non-monotonic dependence on the initial entanglement in many non-ergodic systems, peaking for moderately entangled initial states. To understand this phenomenon, we introduce a conceptual framework that decomposes entanglement growth into two mechanisms: ``build'' and ``move''.  The ``build'' mechanism creates new entanglement, while the ``move'' mechanism redistributes pre-existing entanglement throughout the system. Specifically, we demonstrate that MBL dynamics are ``move-dominated'', exhibiting a quantitative agreement with a random SWAP circuit that serves as a model of pure ``move'' dynamics by uniformly distributing pre-existing entanglement. This implies that MBL acts as a redistributor of a hidden entanglement reservoir quantified by the bipartition-averaged entropy. This ``build-move'' framework offers a unified perspective for classifying diverse physical dynamics, deepening our understanding of entanglement propagation and information processing in quantum many-body systems.

\end{abstract}

\maketitle

\textit{Introduction.---}Entanglement is a cornerstone of modern physics \cite{GUHNE20091, NielsenChuang, RevModPhys.81.865}, and specifically a key diagnostic for non-equilibrium quantum dynamics \cite{doi:10.1126/science.aaf6725, Calabrese_2004, PhysRevA.53.2046, PhysRevB.95.094206, RevModPhys.82.277, RevModPhys.90.035007,ycdh-z8zf}. The standard approach investigates the time evolution of entanglement entropy (EE) from an initial product state, revealing a rich landscape of dynamical behaviors. In thermalizing systems and generic random quantum circuits (RQCs), entanglement grows rapidly, often linearly in time \cite{bianchi_linear_2018, Han2023entanglement, mezei_entanglement_2017, PhysRevA.43.2046, PhysRevB.99.174205, PhysRevE.50.888, PhysRevLett.98.130502, PhysRevLett.109.040502, PhysRevLett.111.127205, PhysRevLett.122.250602, PhysRevX.7.031016, PhysRevX.8.021013, PhysRevX.8.021014, PhysRevX.9.021033, rigol_thermalization_2008, kppn-3272}. In contrast, many-body localized (MBL) systems \cite{BASKO20061126,
PhysRevB.77.064426,
PhysRevB.82.174411,
PhysRevLett.110.067204,
PhysRevLett.111.127201,
PhysRevB.90.174202,
altman2015universal,
campbell_dynamics_2017,
doi:10.1126/science.aaa7432,
doi:10.1126/science.aaf8834,
doi:10.1126/science.aau0818,
FAN2017707,
https://doi.org/10.1002/andp.201600318,
https://doi.org/10.1002/andp.201600332,
LIU20253991,
nandkishore2015many,
PhysRevB.91.140202,
PhysRevB.93.060201,
PhysRevB.94.214206,
PhysRevB.95.045121,
PhysRevB.95.054201,
PhysRevB.95.165136,
PhysRevB.97.214202,
PhysRevB.98.014203,
PhysRevB.99.075162,
PhysRevB.99.184202,
PhysRevB.101.035148,
PhysRevB.105.224203,
PhysRevLett.113.147204,
PhysRevLett.113.217201,
PhysRevLett.114.140401,
PhysRevLett.121.206601,
PhysRevX.5.031032,
RevModPhys.91.021001,
Sierant_2025,
smith_many-body_2016,
Vasseur_2016} exhibit a much slower, logarithmic growth \cite{9jbn-nxwz,
chen2016universallogarithmicscramblingbody,
kim2014localintegralsmotionlogarithmic,
PhysRevB.90.064201,
PhysRevB.95.024202,
PhysRevB.104.214202,
PhysRevLett.109.017202,
PhysRevLett.110.260601,
toniolo2025dynamicsmanybodylocalizedsystems}, while integrable systems saturate to non-thermal values \cite{Calabrese_2005,
Calabrese_2007,
PhysRevX.13.021007,
g1cw-tk7f,
pnf1-r1rm} and non-interacting Anderson localized (AL) systems show severely suppressed entanglement growth \cite{PhysRev.109.1492, billy_direct_2008, Kramer_1993,  PhysRevB.22.4666, PhysRevB.37.325, PhysRevB.95.094204, PhysRevLett.95.206603, PhysRevLett.113.046806, 9kdg-m6yg,RevModPhys.57.287, RevModPhys.80.1355, roati_anderson_2008, TIKHONOV2021168525}. However, this fruitful paradigm inherently conflates two distinct aspects of entanglement dynamics: the initial generation of entangled resources, and the subsequent transport of this entanglement. The observed half-chain entanglement entropy (HCEE) growth is thus always a composite effect, leaving the underlying pattern of entanglement propagation obscured.

To deconstruct this process and isolate the intrinsic entanglement handling capabilities of a given system, we shift the paradigm to study the evolution of initial states that already possess volume-law entanglement, which we term partially thermalized states. These states are thus prepared to contain a rich, pre-existing reservoir of entanglement, providing a non-trivial substrate for the subsequent dynamics. This paradigm shift allows us to move from the question ``how fast is entanglement created?'' to a more profound one: ``how does a system's inherent dynamics process and redistribute pre-existing entanglement?''. This setup facilitates deep probe into the intricate interplay between a system's fundamental character and an existing, complex entanglement structure.

\begin{figure*}[thtbp]
    \centering
    \includegraphics[width=0.88\textwidth]{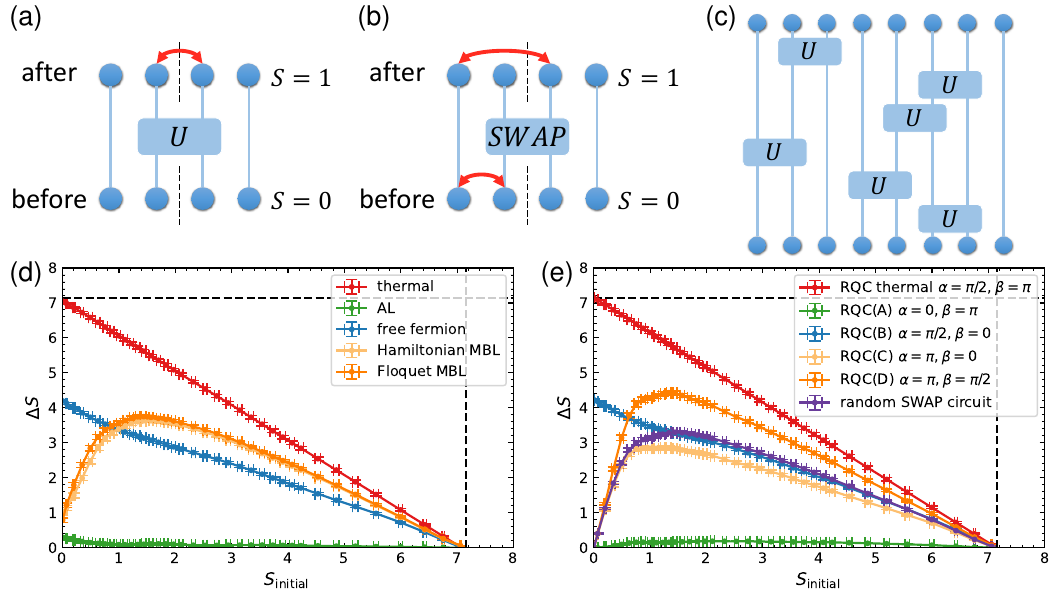} 
    \caption{
    The ``build-move'' framework and the increase of HCEE in various dynamics.
    (a-c) Schematic illustrations of ``build'' mechanism (a), ``move'' mechanism (b), and RQC composed of two-qubit gates $U$ (c). Blue circles represent spins, and pairs of spins linked by red double arrows indicate Bell pairs.
    (d,e) Classification of dynamics based on the HCEE growth: $\Delta S = S_{\text{sat}} - S_{\text{initial}}$, as a function of the initial HCEE, $S_{\text{initial}}$. All data points are for system size $L=16$. $S_{\text{initial}}$ is controlled by varying the preparation time $\tau$.
    (d) Results for Hamiltonian-based dynamics.
    (e) Results for RQC-based dynamics. Note the striking similarity between the MBL curve in (d) and the Random SWAP curve in (e) The two-qubit gates $U$ are defined in \Eq{eq:U_RQC}, with parameters in various classes detailed in Table \ref{tab:1}. For the calculation details of the data in all the figures in the main text, please refer to the Supplemental Material~\cite{SM_reference} Sec. I.
    }
    \label{fig1}
\end{figure*}

In this Letter, we address these questions and uncover a remarkable phenomenon: in MBL dynamics, the HCEE growth exhibits a non-monotonic behavior with its initial value, profoundly differing from thermalized systems (see Fig.~\ref{fig1}(d)). To rationalize the observed behaviors, we introduce a novel conceptual framework that decomposes HCEE growth into two primary mechanisms: ``build'' for creating new entanglement and ``move'' for redistributing pre-existing entanglement, as schematically shown in Fig.~\ref{fig1}(a,b). We demonstrate that this framework not only quantitatively explains our results by identifying MBL as a ``move-dominated'' dynamics, but also provides a unified perspective for classifying diverse quantum dynamics.

\textit{Model and Methods.---}We study the dynamics in a one-dimensional spin-$1/2$ chain of even length $L$ using exact diagonalization. The central quantity of interest is the HCEE measured in bits: $S=-\operatorname{Tr}\left(\hat{\rho}_\text{HC}\log_2\hat{\rho}_\text{HC}\right)$, defined as the von Neumann entropy of the first $L/2$ spins whose reduced density operator is $\hat{\rho}_\text{HC}$. The various physical processes in our study, from initial state preparation to subsequent dynamics, are mainly based on the one-dimensional disordered $XXZ$ Hamiltonian with open boundary conditions:
\begin{equation}
\hat{H}=\sum_{i=1}^{L-1}\left(\hat{S}_{i}^x\hat{S}_{i+1}^x+\hat{S}_{i}^y\hat{S}_{i+1}^y+J_z\hat{S}_{i}^z\hat{S}_{i+1}^z\right)+\sum_{i=1}^{L}h_i\hat{S}_{i}^z.
\label{eq:H_full}
\end{equation}
Here, $\hat{S}_i^{\alpha}$ ($\alpha=x,y,z$) are the spin-$1/2$ operators at site $i$, and we set the anisotropy $J_z=0.5$. The on-site fields $h_i$ are independent random variables drawn uniformly from the interval $\left[-W, W\right]$. This Hamiltonian has a $U(1)$ symmetry which conserves the total magnetization in the $z$-direction, $\hat{S}^z_{\text{tot}} = \sum_{i=1}^L \hat{S}_i^z$. All our calculations are performed in the half-filling sector $\hat{S}^z_{\text{tot}}=0$. The system is in the thermal and MBL phases at $W=0.5$ and $W=5.0$, respectively \cite{PhysRevB.82.174411} (see Supplemental Material (SM)~\cite{SM_reference}\nocite{PhysRevB.75.155111,
doi:10.1126/science.aao1401,
PhysRevLett.52.1,
Mehta2004,
PhysRevLett.110.084101,
zhang2026bondadditivitypersistentgeometric,
PhysRevLett.123.210601,
PhysRevLett.125.070501,
10.21468/SciPostPhys.8.4.067,
PhysRevB.101.094304} Sec. II for details).

\begin{table*}
\caption{\label{tab:table1}Classification of Hamiltonian-based and RQC-based dynamics on the behavior of $\Delta S$ with $S_{\text{initial}}$.}
\begin{ruledtabular}
\begin{tabular}{ccc}
Hamiltonian-based dynamics&RQC-based dynamics& Tendency of $\Delta S$ with $S_{\text{initial}}$\\ \hline
thermalization&$\alpha\in(0,\pi)\cup(\pi,2\pi)$, $\beta\in(0,2\pi)$ &Monotonic decay\\
MBL&$\alpha=\pi$, $\beta =\pi$  (SWAP), $\{0, 2\pi\}$ (class C), other values (class D)&Non-monotonic peak \\
free fermion&$\alpha\in(0,\pi)\cup(\pi,2\pi)$, $\beta=\{0, 2\pi\}$(class B)&Monotonic decay\\
AL&$\alpha=\{0, 2\pi\}$, $\beta\in(0,2\pi)$ (class A)&Negligible\\
\end{tabular}
\end{ruledtabular}
\label{tab:1}
\end{table*}

Our primary focus is on the evolution from initial states with varying degrees of entanglement. We prepare initial states $|\psi(\tau)\rangle$ by evolving a product state $|\psi_0\rangle$ under the thermal Hamiltonian $\hat{H}(W=0.5)$ for a duration $\tau$. $|\psi_0\rangle$ is randomly chosen from the computational basis states within the half-filling sector. The evolution time $\tau$ directly controls the initial entanglement and the level of thermalization as $|\psi(\tau)\rangle=\mathrm{e}^{-\mathrm{i}\tau\hat{H}(W=0.5)}|\psi_0\rangle$.
Subsequently, we examine the dynamics of these prepared states under the following $U(1)$-symmetric dynamical protocols:
\begin{enumerate} 
 \setlength{\itemsep}{0.2pt}
    \setlength{\parskip}{0.2pt}
    \setlength{\parsep}{0.2pt}
    \item \textbf{Thermal Dynamics:} $|\psi(\tau)\rangle$ evolves under the thermal Hamiltonian $\hat{H}(W=0.5)$.
    \item \textbf{Hamiltonian MBL Dynamics:}$|\psi(\tau)\rangle$ evolves under the MBL Hamiltonian $\hat{H}(W=5.0)$.
    \item \textbf{Free Fermion Dynamics:} Evolution under the free fermion Hamiltonian $\hat{H}_\text{XY}$ with $J_z=0$ and $W=0$ in Eq.~\eqref{eq:H_full}.
    \item \textbf{Floquet MBL Dynamics:} A periodic drive is applied using the Floquet operator $\hat{F}=\mathrm{e}^{-\mathrm{i}T_0\hat{H}_0}\mathrm{e}^{-\mathrm{i}T_1\hat{H}_\text{XY}}$ \cite{PhysRevLett.114.140401}. Here, $\hat{H}_0=\sum_{i=1}^{L-1}\hat{S}_{i}^z\hat{S}_{i+1}^z+\sum_{i=1}^{L}h_i\hat{S}_{i}^z$, with disorder strength $W=5.0$, $T_0=1.0$, and $T_1=0.4$.
    \item \textbf{AL Dynamics:} $|\psi(\tau)\rangle$ evolves under the non-interacting AL Hamiltonian, with $J_z = 0$ and $W=5.0$ in Eq.~\eqref{eq:H_full}.
    \item \textbf{RQC protocols:} We employ RQCs as a paradigmatic model for understanding universal aspects of entanglement dynamics \cite{Chen2025subsystem,
    fisher2023random,
    PhysRevA.78.032324,
    PhysRevB.98.205136,
    PhysRevB.99.174205,
    PhysRevB.100.134306,
    PhysRevB.107.L201113,
    PhysRevB.110.064323,
    PhysRevLett.98.130502,
    PhysRevLett.132.240402,
    PhysRevLett.133.140405,
    PhysRevX.8.021013,
    PhysRevX.8.021014,
    PhysRevX.9.031009,
    PhysRevX.11.021040,
wang2024drivencriticaldynamicsmeasurementinduced,10.21468/SciPostPhys.19.5.132}. Our implementation consists of circuits where at each step, a single two-qubit gate $U$ is applied to a randomly selected adjacent pair of spins $(i, i+1)$ from the $L-1$ bonds (Fig.~\ref{fig1}(c)). The two-qubit unitary is given by:
    \begin{equation}
    \hat{U}_{i,i+1}=\mathrm{e}^{-\mathrm{i}\alpha\left(\hat{S}_{i}^x\hat{S}_{i+1}^x+\hat{S}_{i}^y\hat{S}_{i+1}^y\right)}\mathrm{e}^{-\mathrm{i}\beta\hat{S}_{i}^z\hat{S}_{i+1}^z}.
    \label{eq:U_RQC}
    \end{equation}
\end{enumerate}

\textit{Results.---}To analyze the entanglement dynamics under different protocols, we first introduce a conceptual framework that decomposes the growth of HCEE into two fundamental mechanisms: ``build'' and ``move''. As shown in Fig.~\ref{fig1}(a), the ``build'' mechanism creates new entanglement directly. For instance, a two-qubit unitary gate $U$ across the central bond can generate a Bell pair from a product state, increasing the HCEE from $0$ to $1$. The ``move'' mechanism, depicted in Fig.~\ref{fig1}(b), involves the redistribution of pre-existing entanglement. Here, an entangled pair within one subsystem is transported across the central bond, for instance, via a SWAP gate, also raising HCEE from $0$
to $1$. While both mechanisms often occur simultaneously in a generic interacting system, their proportional contributions are highly dependent on the specific dynamics.

To isolate the ``move'' mechanism, we employ the random SWAP circuit as an idealized model, which can be achieved up to a global phase by setting $\alpha=\beta=\pi$ in our RQC protocols. The key insight is that a SWAP operation merely exchanges the state of two spins. This allows us to equivalently view the action of the random SWAP circuit on a fixed bipartition as an evolution of the bipartition itself across a static state. This evolution generates transitions between the $N=\binom{L}{L/2}/2$ distinct bipartitions, establishing an ergodic Markov chain with a uniform stationary distribution (see the proof in SM~\cite{SM_reference} Sec. III). Consequently, at sufficiently large circuit depth, every bipartition becomes equally probable. This directly implies that the steady-state HCEE under the random SWAP circuit corresponds precisely to the initial state's \textit{bipartition-averaged entanglement entropy} (BAEE)---the entanglement entropy averaged across all $N$ possible equal-size bipartitions: $\bar{S}=\frac1N\sum_{\text{A}\in\mathcal{P}}S_\text{A}$,
where $\mathcal{P}$ is the set consisting of subsystems corresponding to all $N$ possible bipartitions and $S_\text{A}$ represents the entropy of the subsystem $\text{A}$ on $\mathcal{P}$. The BAEE provides a more robust measure of the state's overall entanglement, invariant under SWAP dynamics which only redistributes existing entanglement. This model thus establishes a baseline for ``move''-driven dynamics, enabling quantitative comparisons with more complex protocols. For more analysis about the BAEE, see SM~\cite{SM_reference} Sec. VI.

\begin{figure}[!htbp]
    \centering
    \includegraphics[width=0.87\columnwidth]{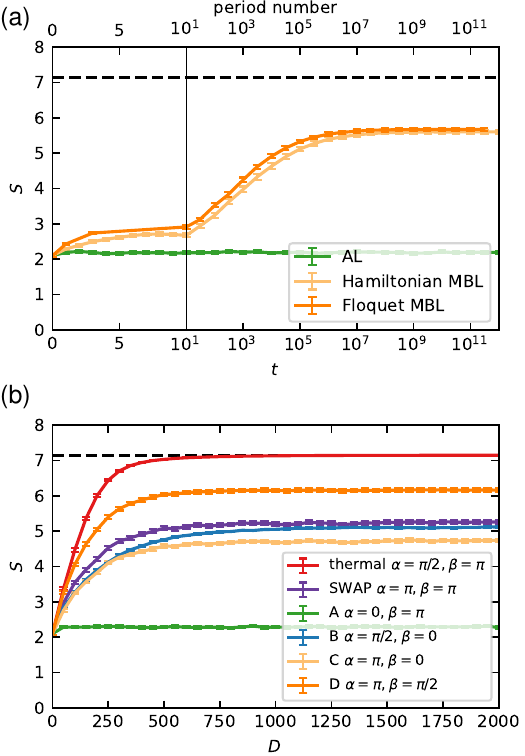}
    \caption{Time evolution of HCEE $S$ with initial state $|\psi(\tau)\rangle$ for $L=16$ and $\tau=4.5$.
    (a) Evolution under localized dynamics. The lower axis shows time $t$ for AL and Hamiltonian MBL evolutions, while the upper axis shows the period number for the Floquet MBL dynamics. To visualize the dynamics across multiple timescales, the horizontal axis uses a hybrid scale of linear for the initial evolution ($t,\text{period number}<10$) and logarithmic for long times.
    (b) Evolution under RQCs dynamics. The horizontal axis represents circuit depth. The specific $(\alpha, \beta)$ values shown in the legend are the parameters of the quantum gates selected from their respective categories.
    }
    \label{fig2}
\end{figure}

\begin{figure}[t]
    \centering
    \includegraphics[width=0.85\columnwidth]{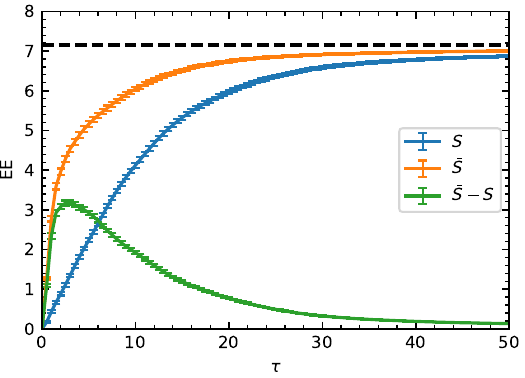}
    \caption{Evolution of BAEE $\bar{S}$, HCEE $S$ and their difference under thermal Hamiltonian $\hat{H}(W=0.5)$ for $L=16$, starting from the product state $|\psi_0\rangle$. The data point corresponding to the maximum value of $\bar{S}-S$ is at $\tau=3.0$ with HCEE $S\approx1.34$, roughly coinciding with the optimal $S_{\text{initial}}$ for the peak $\Delta S$ of move-dominated dynamics in Fig.~\ref{fig1}(d,e).}
    \label{fig3}
\end{figure}

For RQC protocols of $\alpha, \beta \in [0, 2\pi]$, the specific points $(\alpha, \beta)=(0,0),(0,2\pi),(2\pi,0),$ and $(2\pi,2\pi)$ are trivial as the operator $\hat{U}_{i,i+1}$ can be decomposed into a tensor product of single-qubit gates. Generic choices of $(\alpha, \beta)$ are expected to lead to full thermalization, where the HCEE saturates near the Haar measure average value in the half-filling sector \cite{PhysRevB.108.245101,
PhysRevD.100.105010,
PhysRevE.93.052106,
PhysRevLett.71.1291,
PhysRevLett.133.070402} which is marked by dashed lines in Fig.~\ref{fig1}(d,e), Fig.~\ref{fig2} and Fig.~\ref{fig3}. Non-thermalization behavior occurs only for a small set of special parameter values, which we classify into 5 categories based on the distinct saturation values of their HCEE. Within each category, different parameter choices for $(\alpha, \beta)$ lead to dynamics with almost the same saturating HCEE value. The random SWAP circuit mentioned above is one of these 5 categories. The parameter setting for these categories is summarized in Table~\ref{tab:1}.

Fig.~\ref{fig2}(a) displays the localized dynamics for a long time starting from $|\psi(\tau)\rangle$, where the evolution under different non-thermal dynamics reveals distinct behaviors. For both the Hamiltonian MBL and the Floquet MBL protocols, the HCEE exhibits a characteristic slow, logarithmic growth in time. It eventually reaches a saturation value that, while significantly higher than the initial HCEE, remains well below the Haar measure average value in the half-filling sector. In sharp contrast, the dynamics governed by the non-interacting AL Hamiltonian (Fig.~\ref{fig2}(a)) and the RQC(A) protocol (Fig.~\ref{fig2}(b)) both show almost no further increase in HCEE, among which the former confirms that interactions are crucial for moving the entanglement. The other non-thermal RQC protocols, RQC(B-D), all facilitate HCEE growth, but with distinct steady values.

To systematically quantify the relationship between initial entanglement and subsequent entanglement growth, we analyze the total increase of HCEE, $\Delta S = S_{\text{sat}} - S_{\text{initial}}$, as a function of $S_{\text{initial}}$. The results, plotted in Fig.~\ref{fig1}(d,e), reveal that the diverse dynamical protocols can be classified into three classes based on their monotonicity which reflects the capability to generate and transport entanglement. This classification is not only supported by the macroscopic behavior of $\Delta S$, but is also consistent with the microscopic action of the gate for the RQC dynamics.

\textbf{Move-Dominated Dynamics.} The first class of protocols is defined by a predominant ``move'' mechanism, whose fundamental role is to redistribute existing entanglement. As depicted in Fig.~\ref{fig1}(d), the dynamics of MBL clearly exhibit a non-monotonic dependence of $\Delta S$ on $S_{\text{initial}}$. This behavior, which differs significantly from that in thermal or AL dynamics, is a key finding of this Letter. For nearly product states ($S_{\text{initial}} \approx 0$), $\Delta S$ is negligible due to the minimal entanglement available for transport. It then increases to a peak for moderately entangled states before decreasing again as $S_{\text{sat}}$ approaches the Haar measure average value. This distinct non-monotonic behavior serves as a hallmark of the ``move''-dominated mechanism. More importantly, we find the quantitative agreement of the results from MBL and random SWAP circuits, although the time scales for entanglement saturation are distinct for these dynamics. This agreement is highly non-trivial: it implies that despite the complex Hamiltonian origin of MBL, its effective entanglement dynamics approximately reduces to a simple shuffling of bipartitions. This class also encompasses RQC(C) and (D), as illustrated in Fig.~\ref{fig1}(e). This is directly explained by their gate-level action, which functions similarly to a SWAP operator. For example, when an RQC(C) or (D) gate acts on the first two spins of the state $\left\vert\uparrow\uparrow\downarrow\right\rangle+\left\vert\uparrow\downarrow\uparrow\right\rangle$, the transformation yields $\mathrm{e}^{-\mathrm{i}\beta/4}\left\vert\uparrow\uparrow\downarrow\right\rangle-\mathrm{i}\mathrm{e}^{\mathrm{i}\beta/4}\left\vert\downarrow\uparrow\uparrow\right\rangle$. This effectively converts entanglement from between the last two spins to between the first and third spins, thereby realizing the ``move'' operation at the microscopic level.

\textbf{Build and Move Hybrid Dynamics.} The second class of dynamics exhibits a hybrid ``build-and-move'' character. In conventional thermal systems, a significant $\Delta S$ emerges even from initial product states ($S_{\text{initial}} = 0$), signifying ``build'' mechanism that generates entanglement from scratch (Fig.~\ref{fig1}(d)). This entanglement increase, $\Delta S$, diminishes as $S_{\text{initial}}$ rises. Similarly, free fermion systems also reach the equilibrium with weaker ``build'' capabilities due to extensive conserved quantities, resulting in a smaller $\Delta S$. Consistent with this, RQCs with general $\alpha \in (0, \pi) \cup (\pi, 2\pi)$, including both thermal RQC and RQC(B), also show the same monotonically decreasing trend of $\Delta S$, as shown in Fig.~\ref{fig1}(e). The ability of these RQCs to effectively generate entanglement from product states—for instance, transforming $\left\vert\uparrow\downarrow\right\rangle$ into the  entangled state $\mathrm{e}^{\mathrm{i}\beta/4}\left(\cos\frac\alpha2\left\vert\uparrow\downarrow\right\rangle-\mathrm{i}\sin\frac\alpha2\left\vert\downarrow\uparrow\right\rangle\right)$, further corroborates the pervasive  ``build'' mechanism within this class. Specifically, RQC(B) with $\beta=0$ mimics free fermion dynamics. Moreover, with increasing $S_{\text{initial}}$, the ``move'' aspect for these systems also becomes a prominent contributor to the overall entanglement growth.

\textbf{Entanglement-Inert Dynamics.}
Finally, the third class is essentially inert with respect to HCEE growth. Both AL and RQC(A) fall into this class, producing negligible $\Delta S$ across the entire range of $S_{\text{initial}}$ (see Fig.~\ref{fig1}(d,e)). This inert behavior stems from the lack of essential ingredients for either ``build'' or ``move'' contribution. For AL, the non-interacting and localized character of the Hamiltonian prevents entanglement from forming or spreading. For RQC(A), the underlying gate is diagonal in the computational basis, which can only affect relative phases, thereby rendering it incapable of generating or transporting entanglement.

\textit{Discussion and Conclusion.---}In this Letter, we systematically investigate entanglement dynamics across diverse quantum systems, moving beyond the conventional focus on product states to explore initial states that already possess entanglement. Our central finding is a universal three-class classification of entanglement growth. Most notably, we uncover a non-monotonic dependence of HCEE increase on initial entanglement in MBL dynamics (we also considered the quantum sun model~\cite{PhysRevLett.129.060602}, more consistent results are detailed in the SM~\cite{SM_reference}) and specific RQCs---a behavior qualitatively distinct from thermal dynamics. To fundamentally account for these novel results, we introduce the ``build-move'' framework. This framework offers a universal lens through which to understand complex entanglement growth by classifying dynamics based on their capacity to generate new entanglement (``build'') and effectively transport pre-existing entanglement (``move'').

Our study yields another key insight by distinguishing between the entanglement of a single, fixed bipartition (HCEE) and the overall inherent entanglement in the quantum state, which we quantify using the BAEE. Fig.~\ref{fig3} illustrates that, even in a generic thermalized system, the BAEE grows considerably faster than the HCEE during early dynamics. This initial divergence quantifies the rapid ``build''-up of entanglement throughout the system at a local level, effectively establishing an accumulating intra-subsystem entanglement reservoir. This reservoir is central to understanding the initial increase of $\Delta S$ with $S_{\text{initial}}$ in move-dominated dynamics: a larger reservoir provides more entanglement potential to be ``moved'' across the central bond. 
Our results demonstrate that MBL dynamics effectively unlock this potential, converting the hidden reservoir $\bar{S}-S$ into accessible half-chain entropy via the ``move'' mechanism. This picture is supported by a remarkable quantitative consistency: the value of $S_{\text{initial}}$ where the entropy growth $\Delta S$ peaks for MBL dynamics [Fig.~\ref{fig1}(d)] aligns with the maximum difference between BAEE and HCEE. This alignment provides compelling evidence for the dominant ``move'' mechanism in MBL entanglement dynamics, fundamentally recasting MBL as a global redistributor of a ``hidden'' entanglement reservoir.

This perspective underscores the limitation of using a single bipartition's EE to fully capture a system's entanglement structure and highlights the importance of multi-partition entanglement measures, a topic of growing interest \cite{PhysRevB.101.195134,PhysRevB.102.134203,Kolisnyk2026tensorcross}. Our work provides a clear dynamical context for this multi-partition viewpoint. Moreover, the distinct dynamical signatures predicted by our framework present clear targets for experimental verification on near-term quantum simulation platforms. Finally, our results have practical significance for numerical approaches: the saturation entanglement of MBL dynamics which is reached only after exponentially long timescales can be effectively approximated by the BAEE or random SWAP circuits. This provides a basis for utilizing efficient proxies to estimate steady-state properties while bypassing the prohibitive computational cost of simulating the full dynamical evolution.

\textit{Acknowledgement.---}  
C.-Y. Z. and Z.-X. L. are supported by Beijing Natural Science Foundation (No. F251001), the National Natural Science Foundation of China (No. 12347107 and No. 12474146), and the New Cornerstone Investigator Program. S.-X. Z. acknowledges the support from Quantum Science and Technology-National Science and Technology Major Project (No. 2024ZD0301700), the National Natural Science Foundation of China (No. 12574546), and the Chinese Academy of Sciences (No. XDB1680201 and No. YSBR-150).

\bibliographystyle{apsreve}
\let\oldaddcontentsline\addcontentsline
\renewcommand{\addcontentsline}[3]{}
\bibliography{ref}
\let\addcontentsline\oldaddcontentsline

\end{document}


\title{Supplemental Material for ``Entanglement Growth from Entangled States: A Unified Perspective on Entanglement Generation and Transport''}

\author{Chun-Yue Zhang}
\affiliation{Beijing National Laboratory for Condensed Matter Physics $\&$ Institute of Physics, Chinese Academy of Sciences, Beijing 100190, China}
\affiliation{University of Chinese Academy of Sciences, Beijing 100049, China}

\author{Zi-Xiang Li}
\email{zixiangli@iphy.ac.cn}
\affiliation{Beijing National Laboratory for Condensed Matter Physics $\&$ Institute of Physics, Chinese Academy of Sciences, Beijing 100190, China}
\affiliation{University of Chinese Academy of Sciences, Beijing 100049, China}

\author{Shi-Xin Zhang}
\email{shixinzhang@iphy.ac.cn}
\affiliation{Beijing National Laboratory for Condensed Matter Physics $\&$ Institute of Physics, Chinese Academy of Sciences, Beijing 100190, China}

\date{\today}
\maketitle

\onecolumngrid

\renewcommand{\theequation}{S\arabic{equation}}
\renewcommand{\thefigure}{S\arabic{figure}}
\renewcommand{\thetable}{S\arabic{table}}
\renewcommand{\thesection}{\Roman{section}}
\setcounter{equation}{0}
\setcounter{figure}{0}
\setcounter{table}{0}
\setcounter{section}{0}

\renewcommand{\thefigure}{S\arabic{figure}}
\setcounter{figure}{0}
\renewcommand{\theequation}{S\arabic{equation}}
\setcounter{equation}{0}
\renewcommand{\thesection}{\Roman{section}}
\setcounter{section}{0}
\setcounter{secnumdepth}{4}

\addtocontents{toc}{\protect\setcounter{tocdepth}{0}}
{
\tableofcontents
}

\section{Numerical Details in the Main Text}\label{sec:numerical_details}

The data presented in Fig.~1(d,e), Fig.~2 and Fig.~3 are all averaged over $72$ independent runs. Each run uses one $|\psi_0\rangle$ sampled from the computational basis states within the half-filling sector and six sets of independently sampled disorders with strengths for corresponding Hamiltonian-based evolutions:

(i) one set of disorders for the thermal Hamiltonian to prepare initial state $|\psi(\tau)\rangle$ in Fig.~1(d,e) and Fig.~2;

(ii) four sets of disorders for thermal,  Hamiltonian MBL, Floquet MBL and AL dynamics to quench $|\psi(\tau)\rangle$ in Fig.~1(d) and Fig.~2(a);

(iii) one set of disorders for the thermal Hamiltonian to evolve $|\psi_0\rangle$ in Fig.~3.

For the RQC protocols, the results for each run are further averaged over $5$ independent circuit realizations (the bond to apply gate $U$ at each depth).

The values of $\tau$ used to generate $S_{\text{initial}}$ of the data points in Fig.~1(d,e) are as follows: 

$0.0$, $0.25$, $0.5$, $0.75$, $1.0$, $1.25$, $1.5$, $1.75$, $2.0$, $2.25$, $2.5$, $2.75$, $3.0$, $3.3$, $3.6$, $3.9$, $4.2$, $4.5$, $5.0$, $5.5$, $6.0$, $6.5$, $7.0$, $7.5$, $8.0$, $8.5$, $9.0$, $9.5$, $10.0$, $11.0$, $12.2$, $13.7$, $15.7$, $19.0$, $24.0$, $32.0$, $500.0$.

The values of $S_{\text{sat}}$ in Fig.~1(d,e) are determined as follows: for thermal, Hamiltonian MBL and AL dynamics, it is the value at $t=10^{12}$; for Floquet MBL, it is the value at the $3 \times 10^{11}$-th period; for the random SWAP circuit, it is the BAEE of the initial state $|\psi(\tau)\rangle$; for free fermion dynamics and RQC protocols except for random SWAP circuit, it is the average value over 100 late-time/depth snapshots (free fermion: evolution time $t=201.0, 202.0, 203.0, \cdots, 300.0$; thermal and A-D classes of RQCs: depth $D=1901, 1902, 1903, \cdots, 2000$).

\section{The Level Spacing Ratios Distribution Analysis for Thermalization and MBL}\label{sec:sm_level_spacing}

The distribution of level spacing ratios serves as a powerful probe to distinguish between thermal and MBL phases, a method rooted in random matrix theory \cite{PhysRevB.75.155111,PhysRevB.82.174411,doi:10.1126/science.aao1401,PhysRevLett.52.1,Mehta2004}. We employ this diagnostic for the disordered $XXZ$ Hamiltonian:
\begin{equation}
\hat{H}=\sum_{i=1}^{L-1}\left(\hat{S}_{i}^x\hat{S}_{i+1}^x+\hat{S}_{i}^y\hat{S}_{i+1}^y+J_z\hat{S}_{i}^z\hat{S}_{i+1}^z\right)+\sum_{i=1}^{L}h_i\hat{S}_{i}^z,
\label{eqS1}
\end{equation}
where $J_z=0.5$ and the on-site fields $h_i$ are independent random variables uniformly distributed in $[-W, W]$. It is well-established that this model undergoes a phase transition from a thermal phase at weak disorder to a MBL phase at strong disorder.

Due to the $U(1)$ symmetry of $\hat{H}$, the total magnetization $\hat{S}^z_{\text{tot}}=\sum_{i=1}^{L}\hat{S}_{i}^z$ is a conserved quantity. Our analysis is performed via exact diagonalization within the largest symmetry sector, $\hat{S}^z_{\text{tot}}=0$. To minimize finite-size effects and probe the bulk spectral properties, we sorted all eigenenergies within this sector in order and selected the middle one-third for our analysis. This selection process is consistent with the approach in Ref.~\cite{PhysRevB.82.174411}.

\begin{figure}[!htbp]
    \centering
    \includegraphics[width=1.0\columnwidth]{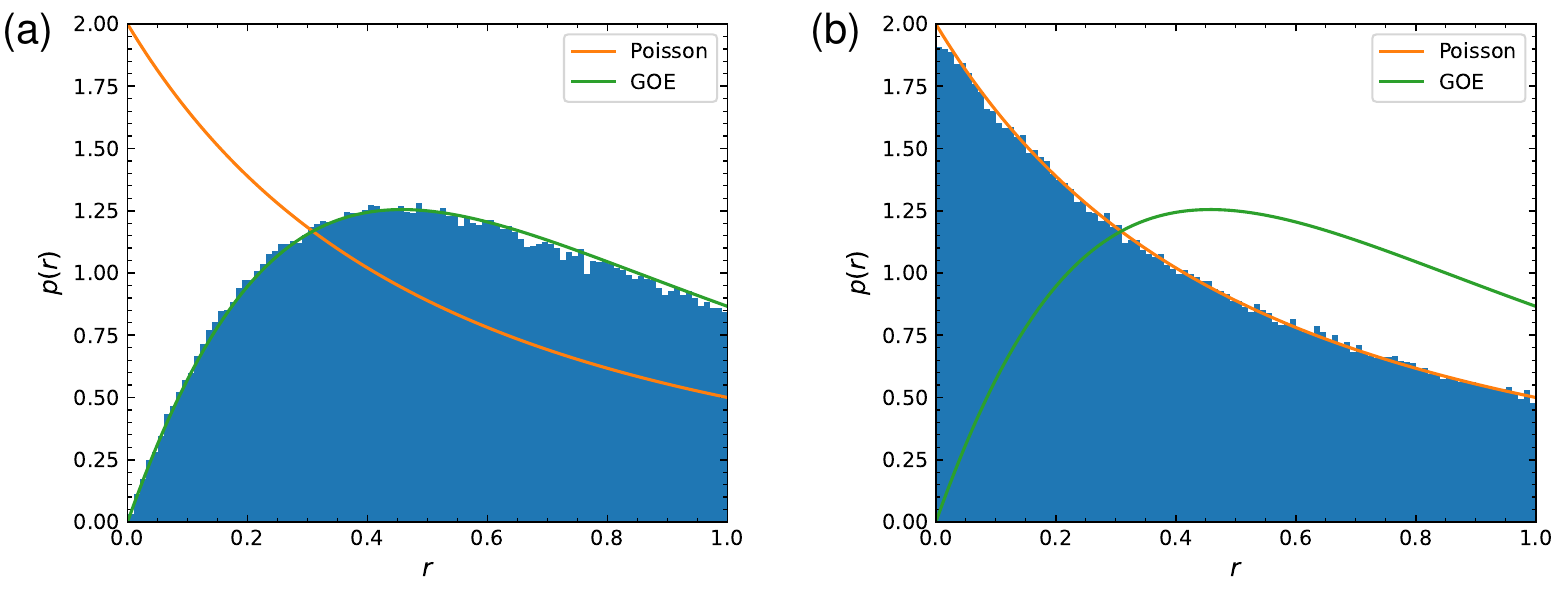}
    \caption{Level spacing ratios distribution $p(r)$ of the Hamiltonian $\hat{H}$ (Eq.~\eqref{eqS1}) for $L=16$. For each panel, the distribution is computed from the middle one-third of the eigenenergies in the half-filling sector, with data compiled from 100 independent disorder realizations for the corresponding disorder strength $W$. Solid lines represent the theoretical predictions.
    (a) Thermal phase ($W=0.5$). The distribution clearly follows the Wigner-Dyson form for the GOE. The numerically obtained average value of $\left\lbrace r_n\right\rbrace$ is about $0.531$, is in excellent agreement with the GOE prediction $\bar{r}=4-2\sqrt3\approx0.536$ \cite{PhysRevLett.110.084101}.
    (b) MBL phase ($W=5.0$). The distribution is well-described by Poisson statistics. The numerical average value of $\left\lbrace r_n\right\rbrace$ is about $0.388$, closely matches the theoretical value for a Poisson distribution, $\bar{r}=2\ln2-1\approx0.386$ \cite{PhysRevLett.110.084101}.}
    \label{figs1}
\end{figure}

The level spacing ratio for adjacent energy levels $\left\lbrace E_n\right\rbrace$ labeled in ascending order from smallest to largest is defined as:
\begin{equation}
r_n=\min\left(\frac{\delta_{n+1}}{\delta_{n}},\frac{\delta_{n}}{\delta_{n+1}}\right),
\end{equation}
where $\delta_n = E_{n+1} - E_n$ represents the spacing between consecutive eigenenergies. We then construct histograms of the $r_n$ distribution, $p(r)$ (see Fig.~\ref{figs1}). In the thermalized phase, the distribution is expected to follow the Wigner-Dyson distribution, specifically the Gaussian Orthogonal Ensemble (GOE), while in the MBL phase, it should approach a Poisson distribution. The analytical expressions for $p(r)$ and the corresponding mean values of $r$ (denoted by $\bar{r}$) for both Poisson and GOE statistics can be found in Ref.~\cite{PhysRevLett.110.084101}. These theoretical predictions serve as benchmarks for identifying the phase of the system based on the observed level spacing statistics. Fig.~\ref{figs1} shows that $\hat{H}(W=0.5)$ and $\hat{H}(W=5.0)$ indeed lie in thermal phase and MBL phase, respectively.

\section{The Stationary Distribution of the Bipartitions under the Random SWAP Circuit}\label{sec:sm_bipartition_Markov}

Here, we provide an analytical proof that the Markov chain describing the evolution of bipartitions under the random SWAP circuit is ergodic and has a uniform stationary distribution. This Markov chain is time-homogeneous. Its state space consists of the $N=\binom{L}{L/2}/2$ distinct ways to partition the spin chain of even length $L$ into two subsystems of size $L/2$.

We first establish the ergodicity of this Markov chain. For such a chain over a finite state space, a fundamental theorem states that ergodicity is equivalent to the chain being both irreducible and aperiodic. We now demonstrate these two properties.

The chain is irreducible because the set of adjacent SWAP gates generates the entire symmetric group of permutations. This means that any order of spins can be transformed into any other through a sequence of SWAP operations, ensuring that any bipartition is reachable from any other.

The chain is also aperiodic. For $L\geqslant4$, there always exists at least one SWAP gate that acts on two spins within the same subsystem of a given bipartition (e.g., the standard half-chain bipartition). Such an operation leaves the bipartition unchanged, resulting in a non-zero self-transition probability, $P_{ii}>0$. The existence of such a 1-step return path means that the set of all possible numbers of steps to return to state $i$ contains the integer 1, which confirms that the chain is aperiodic.

Having established both irreducibility and aperiodicity, we conclude that this Markov chain is ergodic.

Next, we show that the stationary distribution $w$ is uniform by invoking the detailed balance condition, $w_i P_{ij} = w_j P_{ji}$. For a uniform distribution, where $w_i$ is constant for all states $i$, this condition simplifies to $P_{ij} = P_{ji}$. The transition probability from bipartition $i$ to $j$ is $P_{ij} = k_{ij}/(L-1)$, where $k_{ij}$ is the number of SWAP gates that transform $i$ to $j$. Since the SWAP gate is its own inverse, the exact same set of gates that transforms $i$ to $j$ also transforms $j$ to $i$. Thus, $k_{ij} = k_{ji}$, which implies $P_{ij} = P_{ji}$. The detailed balance condition is therefore satisfied for a uniform distribution.

Finally, the ergodic theorem guarantees that an ergodic Markov chain converges to a unique stationary distribution irrespective of the initial distribution. We thus conclude that the random SWAP circuit drives the system towards a uniform stationary distribution over all possible equal-sized bipartitions.

\section{MBL Dynamics from Product and Bell-like States}

To provide additional context for our main findings and to further probe the nature of the ``move'' mechanism in MBL systems, we investigate the MBL dynamics starting from two fundamental classes of initial states: product states and ``Bell-like'' entangled states. This analysis directly illustrates the significant difference between ``building'' entanglement from scratch and ``moving'' a pre-existing entanglement reservoir.

Fig.~\ref{prl_reply_product_valence_bond_MBL_evolution} shows the full time evolution of the HCEE under Hamiltonian MBL dynamics for a variety of these initial states. The results reveal two distinct behaviors:

\begin{figure}[!htbp]
    \centering
    \includegraphics[width=0.95\columnwidth]{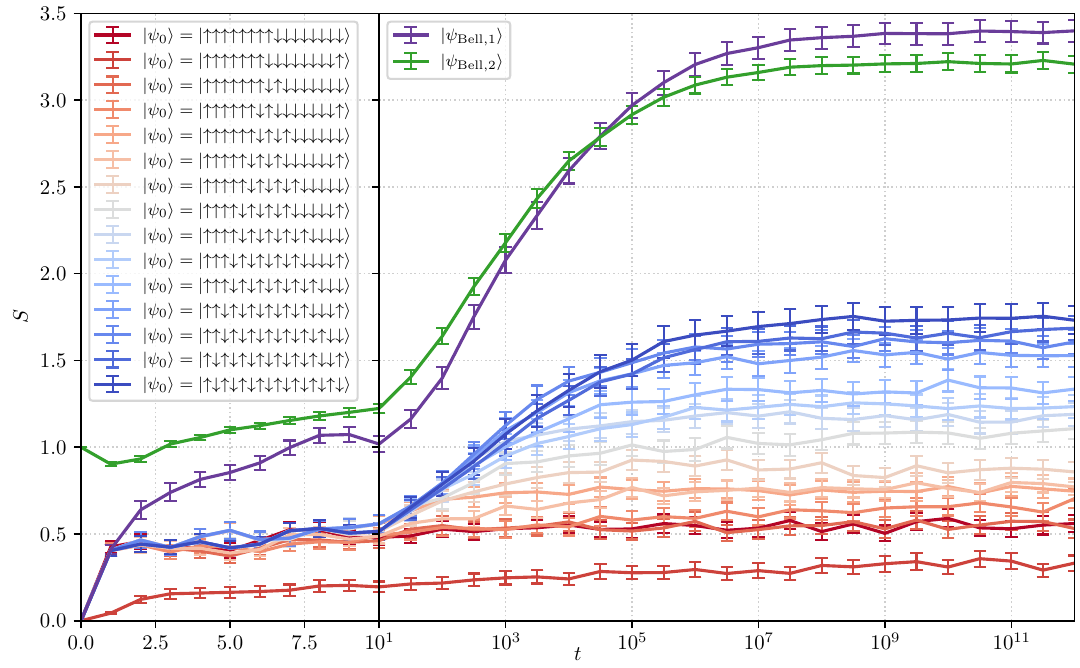}
    \caption{Time evolution of HCEE $S$ with distinct initial states for Hamiltonian MBL dynamics  (governed by $\hat{H}(W=5.0)$, as described in the main text). All data points are for system size $L=16$ and are averaged over 72 independent runs. To visualize the dynamics across multiple timescales, the horizontal axis uses a hybrid scale: it is linear for the initial evolution ($t<10$) and becomes logarithmic for long times.}
    \label{prl_reply_product_valence_bond_MBL_evolution}
\end{figure}

First, the evolution from various product states $|\psi_0\rangle$ with different numbers of domain walls (from 1 to $L-1$, shown in red to blue) all exhibit the characteristic slow, logarithmic growth of entanglement. They eventually saturate to low, non-thermal values, consistent with the established phenomenology of MBL and its weak entanglement ``building'' capacity.

Second, we consider two specific Bell-like initial states, constructed from singlet pairs:
\begin{equation}
\begin{aligned}
|\psi_\text{Bell,1}\rangle=\prod_{i=1}^{L/2}\otimes\left[\frac{1}{\sqrt{2}}\left(\left|\uparrow\downarrow\right\rangle-\left|\downarrow\uparrow\right\rangle\right)\right],
\end{aligned}
\label{eq:psi_Bell_1}
\end{equation}
\begin{equation}
\begin{aligned}
|\psi_\text{Bell,2}\rangle=\left|\uparrow\right\rangle\otimes\left\lbrace\prod_{i=1}^{L/2-1}\otimes\left[\frac{1}{\sqrt{2}}\left(\left|\uparrow\downarrow\right\rangle-\left|\downarrow\uparrow\right\rangle\right)\right]\right\rbrace\otimes\left|\downarrow\right\rangle.
\end{aligned}
\label{eq:psi_Bell_2}
\end{equation}
The state $|\psi_\text{Bell,1}\rangle$ is a tensor product of $L/2$ adjacent singlet pairs, while $|\psi_\text{Bell,2}\rangle$ contains $L/2-1$ singlet pairs. For the system size $L=16$, their initial HCEE values are exactly $S=0$ and $S=1$, respectively.

As shown by the purple and green curves in Fig.~\ref{prl_reply_product_valence_bond_MBL_evolution}, the dynamics starting from these Bell-like states are significantly different. They exhibit a faster and larger entanglement growth compared to the product states. This is a direct and powerful manifestation of the ``move'' mechanism. While product states slowly ``build'' entanglement from scratch due to the very limited build capacity in MBL systems, the Bell-like states start with a rich reservoir of short-range entanglement. The MBL dynamics then efficiently ``move'' this pre-existing entanglement across the chain, leading to a more rapid increase in the HCEE. In particular, the saturation value of HCEE starting from $|\psi_\text{Bell,1}\rangle$ is greater than that of $|\psi_\text{Bell,2}\rangle$. This is because $|\psi_\text{Bell,1}\rangle$ has one more ``bell pair'' compared to $|\psi_\text{Bell,2}\rangle$, and thus has more initial entanglement for subsequent MBL dynamics to move. All these behaviors strongly support our central claim that MBL acts as an effective redistributor of existing entanglement with very limited entanglement building power.

\section{Other Types of Entangled Initial State and the Corresponding Results}\label{sec:other_entangled_initial_state}

To further probe the ``build-move'' framework and provide more targeted evidence for the ``move-dominated'' nature of MBL dynamics, we investigate the system's evolution starting from two distinct types of entangled states: ``locally entangled'' states and the eigenstates of the thermal Hamiltonian. In this section, the setups of Hamiltonian MBL and Floquet MBL dynamics are the same as those in the main text. Dashed lines in Fig.~\ref{figs2}, Fig.~\ref{figs3}, Fig.~\ref{figs4} and Fig.~\ref{figs5} still mark the Haar measure average HCEE in the half-filling sector.

\subsection{Locally Entangled States}

To create an ideal scenario for testing the ``move'' mechanism, we prepare initial states that possess significant entanglement within each half of the system but have strictly zero HCEE by construction. This provides a direct measure of the system's capacity to ``move'' entanglement.

These ``locally entangled'' states, denoted by $\left\vert\psi'\left(\tau\right)\right\rangle$, are prepared by evolving a sampled $\left\vert\psi_0\right\rangle$ for a duration $\tau$ under a modified Hamiltonian, $\hat{H}_\text{local}$, where the interaction across the central bond $(L/2,L/2+1)$ is explicitly removed:
\begin{equation}
\hat{H}_\text{local}=\hat{H}(W=0.5)-\left(\hat{S}_{L/2}^x\hat{S}_{L/2+1}^x+\hat{S}_{L/2}^y\hat{S}_{L/2+1}^y+J_z\hat{S}_{L/2}^z\hat{S}_{L/2+1}^z\right).
\label{eq:H_local}
\end{equation}
The definitions of $\left\vert\psi_0\right\rangle$ and $\hat{H}$ are the same as those in the main text. We still set $J_z=0.5$ and $\hat{H}(W=0.5)$ is thermal as mentioned before. Consequently,
\begin{equation}
|\psi'(\tau)\rangle =\mathrm{e}^{-\mathrm{i}\tau\hat{H}_\text{local}}|\psi_0\rangle.
\label{eq:psi_tau_prime}
\end{equation}
By design, the HCEE of $\left\vert\psi'\left(\tau\right)\right\rangle$ is zero, while the amount of entanglement within each subsystem serves as a ``reservoir'' that can be tuned by the preparation time $\tau$.

\begin{figure}[!htbp]
    \centering
    \includegraphics[width=0.45\columnwidth]{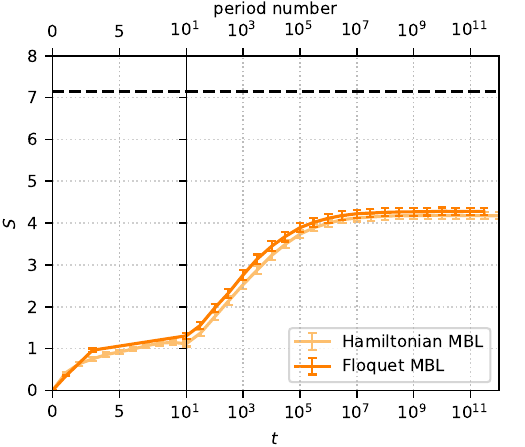}
    \caption{Time evolution of HCEE $S$ with initial state $|\psi'(\tau)\rangle$ for $L=16$ and $\tau=3.0$. The lower axis shows time $t$ for the Hamiltonian MBL evolution, while the upper axis shows the period number for the Floquet MBL dynamics. To visualize the dynamics across multiple timescales, the horizontal axis uses a hybrid scale: it is linear for the initial evolution ($t,\text{period number}<10$) and becomes logarithmic for long times.}
    \label{figs2}
\end{figure}

Fig.~\ref{figs2} shows the MBL evolution of the HCEE, starting from $\left\vert\psi'\left(\tau\right)\right\rangle$. As expected, the HCEE increases from $0$, still exhibits logarithmic growth over time and eventually saturates at significantly increased but non-thermal values.

\begin{figure}[!htbp]
    \centering
    \includegraphics[width=0.58\columnwidth]{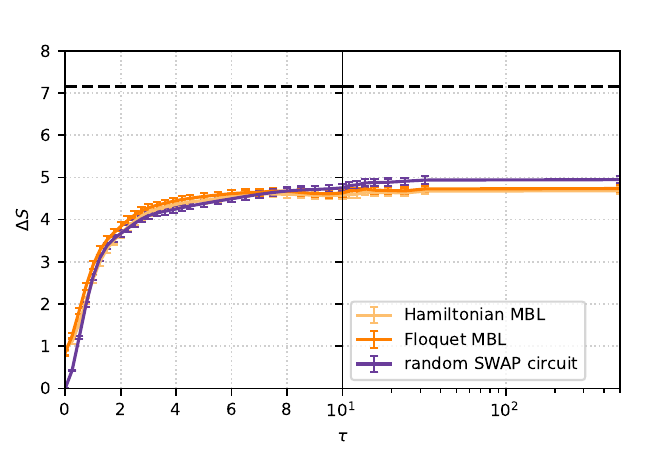}
    \caption{The total growth of HCEE as a function of the preparation time $\tau$ with initial state $|\psi'(\tau)\rangle$ for $L=16$. In order to achieve good visualization of the data, the horizontal axis uses a hybrid scale: it is linear for the small $\tau$ ($\tau<10$) and becomes logarithmic for large $\tau$.
    }
    \label{figs3}
\end{figure}

Fig.~\ref{figs3} plots the total HCEE growth. $\Delta S = S_{\text{sat}}-S_{\text{initial}}=S_{\text{sat}}$ as $S_{\text{initial}}=0$. The values of $\tau$ for the data points and the ways to determine the values of $S_{\text{sat}}$ are the same as those for Fig.~1(d,e) in the main text, both of which have been explained in detail in Sec.~\ref{sec:numerical_details}.

The data presented in Fig.~\ref{figs2} and Fig.~\ref{figs3} are all averaged over $72$ independent runs. Each run uses one sampled $|\psi_0\rangle$ and three sets of independently sampled disorders with strengths for corresponding Hamiltonian-based evolutions: one set of disorders for $\hat{H}_\text{local}$ to prepare $|\psi'(\tau)\rangle$ and two sets of disorders for Hamiltonian MBL and Floquet MBL dynamics to quench $|\psi'(\tau)\rangle$.

Since the lines of 2 kinds of MBL dynamics are very close to the line of the random SWAP circuit for almost the whole range of $\tau$, Fig.~\ref{figs3} once again demonstrates that MBL is a kind of move-dominated dynamics for the growth of the HCEE.

\subsection{Eigenstates of Thermal Hamiltonian}

\begin{figure}[!htbp]
    \centering
    \includegraphics[width=0.45\columnwidth]{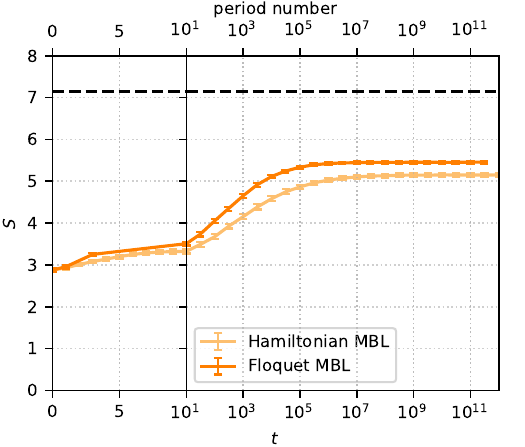}
    \caption{Time evolution of HCEE $S$ starting from the eigenstate whose energy ranks 29th from the lowest to the highest in the half-filling sector of $\hat{H}(W=0.5)$ for $L=16$. The lower axis shows time $t$ for the Hamiltonian MBL evolution, while the upper axis shows the period number for the Floquet MBL dynamics. To visualize the dynamics across multiple timescales, the horizontal axis uses a hybrid scale: it is linear for the initial evolution ($t,\text{period number}<10$) and becomes logarithmic for long times.}
    \label{figs4}
\end{figure}

We now consider the MBL dynamics originating from the eigenstates of the thermal Hamiltonian $\hat{H}(W=0.5)$. These eigenstates span a wide range of HCEE values, providing a natural ensemble of initial states with varying degrees of entanglement. Fig.~\ref{figs4} shows the results for the eigenstate whose energy ranks 29th from the lowest to the highest in the half-filling sector. The logarithmic growth over time and the non-thermal saturation value of HCEE have been maintained.

\begin{figure}[!htbp]
    \centering
    \includegraphics[width=0.58\columnwidth]{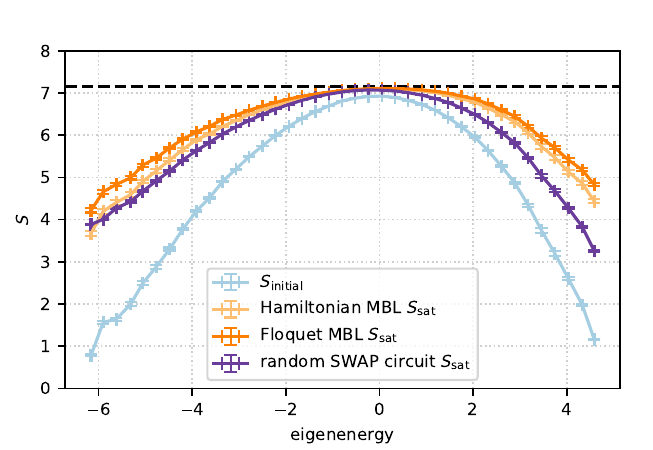}
    \caption{The initial HCEE $S_{\rm initial}$ and saturated HCEE $S_{\rm sat}$ as a function of the eigenenergies. Initial states are the eigenstates in the half-filling sector of $\hat{H}(W=0.5)$ for $L=16$.}
    \label{figs5}
\end{figure}

Fig.~\ref{figs5} plots the initial and saturated HCEEs for many eigenstates. The dimension of the half-filling sector for $L=16$ is $12870$. We sorted the eigenstates within it in ascending order of energy and selected the 1st, 2nd, 4th, 8th, 15th, 29th, 52nd, 87th, 142nd, 222nd, 337th, 494th, 704th, 978th, 1324th, 1750th, 2259th, 2855th, 3533rd, 4283rd, 5095th, 5950th, 6826th, 7700th, 8548th, 9348th, 10079th, 10727th, 11280th, 11737th, 12098th, 12371st, 12568th, 12701st, 12785th, 12833rd, 12857th, 12867th and 12870th ones to calculate the corresponding data. The ways to determine the values of $S_{\text{sat}}$ are also the same as those for Fig.~1(d,e) in the main text, which have been explained in detail in Sec.~\ref{sec:numerical_details}.

The data presented in Fig.~\ref{figs4} and Fig.~\ref{figs5} are all averaged over $72$ independent runs. Each run uses three sets of independently sampled disorders: one set of weak disorders for the thermal Hamiltonian to provide eigenstates and two sets of strong disorders for the quench dynamics of Hamiltonian MBL and Floquet MBL.

We can see that all four lines of $S_{\text{initial}}$ and $S_{\text{sat}}$ in Fig.~\ref{figs5} show a trend of first increasing and then decreasing. They reach their maximum values when eigenenergy is approximately equal to 0. By comparing the $S_{\text{sat}}$ lines of two kinds of MBL dynamics to the $S_{\text{sat}}$ line of the random SWAP circuit, we can also confirm that MBL is a kind of move-dominated dynamics for the growth of the HCEE.

\section{More Analysis about BAEE}

In the main text, our analysis primarily uses the HCEE to characterize the initial states and to quantify entanglement growth. This choice is motivated by the fact that HCEE is the most standard probe of entanglement in studies of one-dimensional quantum many-body systems. Our narrative begins with the discovery of a novel, non-monotonic behavior in this standard quantity, which we then explain using the BAEE, $\bar{S}$ as a tool to reveal the underlying ``entanglement reservoir''.

Here, we present a more detailed analysis, which further strengthens our ``build-move'' framework. We investigate the growth of BAEE ($\Delta \bar{S} = \bar{S}_{\text{sat}} - \bar{S}_{\text{initial}}$) alongside the growth of HCEE ($\Delta S = S_{\text{sat}} - S_{\text{initial}}$) for different dynamical regimes. Dashed lines in Fig.~\ref{prl_reply_AL_Delta_HCEE_Delta_BAEE}, Fig.~\ref{prl_reply_RQC_A_Delta_HCEE_Delta_BAEE}, Fig.~\ref{prl_reply_MBL_Delta_HCEE_Delta_BAEE} and Fig.~\ref{prl_reply_MBL_Delta_HCEE_vs_initial_BAEE} still mark the Haar measure average HCEE in the half-filling sector.

\subsection{Entanglement-Inert Dynamics}

\begin{figure}[!htbp]
    \centering
    \includegraphics[width=0.58\columnwidth]{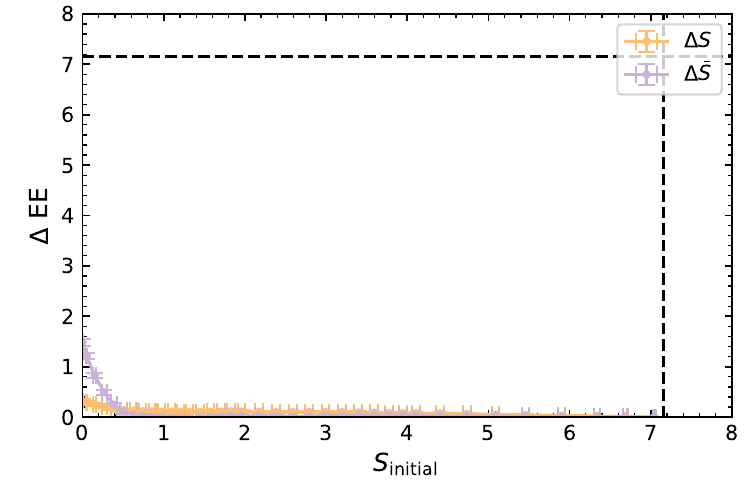}
    \caption{The HCEE growth and BAEE growth as a function of the initial HCEE $S_{\text{initial}}$ for Anderson localized (AL) dynamics according to the setup described in the main text. All data points are for system size $L=16$. The evolution time of saturated entanglement entropy is $t=10^{12}$. The data are averaged over 72 independent runs.}
    \label{prl_reply_AL_Delta_HCEE_Delta_BAEE}
\end{figure}

\begin{figure}[!htbp]
    \centering
    \includegraphics[width=0.58\columnwidth]{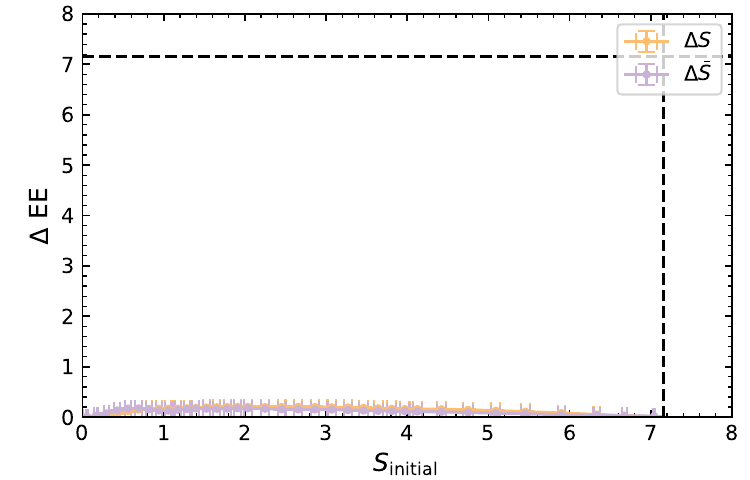}
    \caption{The HCEE growth and BAEE growth as a function of the initial HCEE $S_{\text{initial}}$ for RQC(A) dynamics according to the setup described in the main text. All data points are for system size $L=16$. The evolution depth of saturated entanglement entropy is $D=500$. The data are averaged over 64 independent runs. The evolution process of the random circuit for each run are further averaged over 5 independent circuit realizations.}
    \label{prl_reply_RQC_A_Delta_HCEE_Delta_BAEE}
\end{figure}

For entanglement-inert dynamics, we expect both the ``build'' and ``move'' mechanisms to be suppressed. Our numerical results, shown in Fig.~\ref{prl_reply_AL_Delta_HCEE_Delta_BAEE} and Fig.~\ref{prl_reply_RQC_A_Delta_HCEE_Delta_BAEE}, confirm this expectation. For both AL and RQC(A) dynamics, the growth of both HCEE and BAEE is negligible across the entire range of initial entanglement. This provides a sharp, quantitative confirmation that these systems are truly ``inert'': they neither generate new entanglement throughout the system (near-zero $\Delta \bar{S}$) nor transport it across the half-chain bipartition (near-zero $\Delta S$).

\subsection{Hamiltonian MBL Dynamics}

\begin{figure}[!htbp]
    \centering
    \includegraphics[width=0.58\columnwidth]{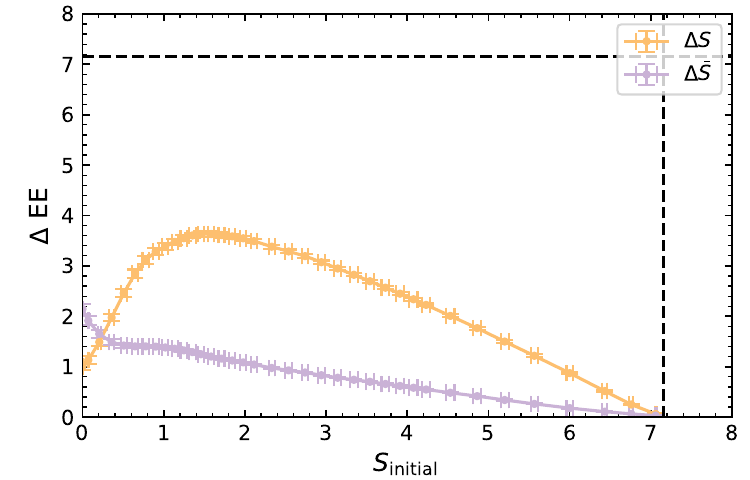}
    \caption{The HCEE growth and BAEE growth as a function of the initial HCEE $S_{\text{initial}}$ for Hamiltonian MBL dynamics according to the setup described in the main text. All data points are for system size $L=16$. The evolution time of saturated entanglement entropy is $t=10^{12}$. The data are averaged over 63 independent runs.}
    \label{prl_reply_MBL_Delta_HCEE_Delta_BAEE}
\end{figure}

The comparison between $\Delta S$ and $\Delta \bar{S}$ for Hamiltonian MBL dynamics provides a particularly deep insight into its ``build'' and ``move'' mechanisms. As shown in Fig.~\ref{prl_reply_MBL_Delta_HCEE_Delta_BAEE}, while the HCEE growth $\Delta S$ exhibits its characteristic non-monotonic peak, the BAEE growth $\Delta \bar{S}$—which quantifies the net ``build'' component—is significantly small and decays almost monotonically, consistent with our claim that MBL dynamics is ``move-dominated''.

For very small $S_{\text{initial}}$, we observe that $\Delta S < \Delta \bar{S}$. This initial stage has very limited pre-existing entanglement reservoir to ``move'' and is thus governed by the weak ``build'' mechanism. As we detail in our related work \cite{zhang2026bondadditivitypersistentgeometric}, MBL dynamics generate a highly structured entanglement pattern dominated by short-range correlations. This makes HCEE smaller than the EEs under other equal-sized bipartitions, thus HCEE is less than BAEE. Our central claim is further supported in the regime of moderate to large degree of entanglement of the initial state. As $S_{\text{initial}}$ increases, we find $\Delta S > \Delta \bar{S}$ and $\Delta \bar{S}$ is very small. In this case, the entanglement reservoir becomes large enough and it enables the ``move'' mechanism of MBL to be fully exerted. This provides quantitative evidence for our central finding that MBL dynamics are move-dominated.

\begin{figure}[!htbp]
    \centering
    \includegraphics[width=0.58\columnwidth]{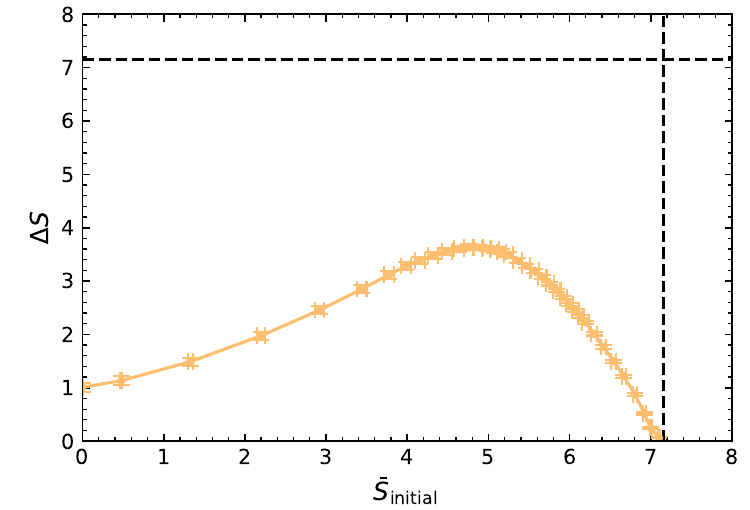}
    \caption{The HCEE growth as a function of the initial BAEE $\bar{S}_{\text{initial}}$ for Hamiltonian MBL dynamics according to the setup described in the main text. All data points are for system size $L=16$. The evolution time of saturated entanglement entropy is $t=10^{12}$. The data are averaged over 63 independent runs.}
    \label{prl_reply_MBL_Delta_HCEE_vs_initial_BAEE}
\end{figure}

Finally, to address the possibility of parameterizing the initial entanglement by the BAEE, we plot the HCEE growth $\Delta S$ as a function of $\bar{S}_{\text{initial}}$ in Fig.~\ref{prl_reply_MBL_Delta_HCEE_vs_initial_BAEE}. This change of axis preserves the non-monotonic nature of the HCEE growth, merely shifting the peak's location. This non-monotonic behavior is still the hallmark of move-dominated dynamics, independent of how the initial state's entanglement is parameterized.

\section{Robustness and Universality of Non-Monotonic Entanglement Growth in MBL Systems}

A central finding of our work is the non-monotonic growth of HCEE as a function of initial entanglement in MBL systems, which we identify as a hallmark of ``move-dominated'' dynamics. To demonstrate the robustness and universality of this phenomenon, we present here a series of additional numerical results. We show that this non-monotonic behavior is not an artifact of specific model parameters (such as disorder strength $W$ or anisotropy $J_z$) and that it persists in a different, well-established model of MBL.

\subsection{Robustness against Disorder Strength}

\begin{figure}[!htbp]
    \centering
    \includegraphics[width=0.58\columnwidth]{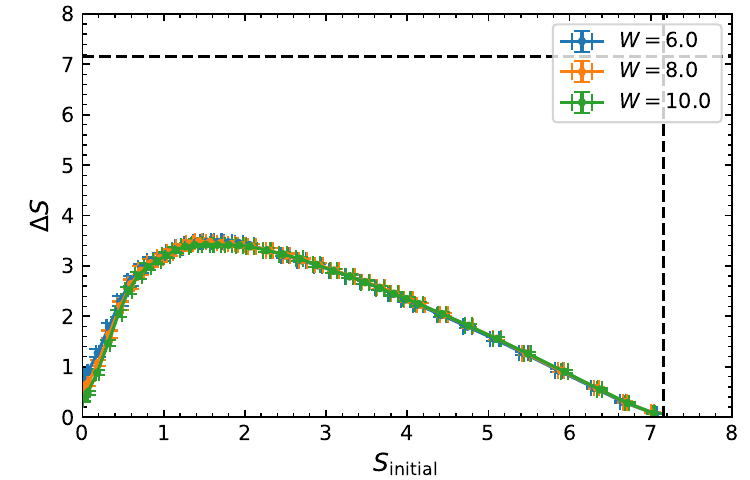}
    \caption{The HCEE growth as a function of the initial HCEE $S_{\text{initial}}$ for Hamiltonian MBL dynamics according to the setup described in the main text. The disorder strength of the MBL dynamics was increased to $W=6.0$, $W=8.0$ and $W=10.0$. All data points are for system size $L=16$. The evolution time of saturated entanglement entropy is $t=10^{12}$. The data for $W=6.0$ are averaged over 68 independent runs. The data for $W=8.0$ and $W=10.0$ are averaged over 72 independent runs. Dashed lines mark the Haar measure average HCEE in the half-filling sector.}
    \label{prl_reply_higher_W_MBL_Delta_HCEE}
\end{figure}

To test the dependence of our results on the disorder strength, we have performed simulations for the Hamiltonian MBL dynamics across a range of strong $W$ values, $6.0$, $8.0$ and $10.0$, in addition to the $W=5.0$ case presented in the main text. The resulting HCEE growth as a function of initial entanglement is shown in Fig.~\ref{prl_reply_higher_W_MBL_Delta_HCEE}. The data reveal that the characteristic non-monotonic profile of $\Delta S$ is stable against variations in disorder strength of MBL dynamics. The curves for different values of $W$ are nearly indistinguishable, demonstrating that the ``move-dominated'' behavior is a generic feature deep within the MBL phase.

\subsection{Robustness against Anisotropy Parameter}

\begin{figure}[!htbp]
    \centering
    \includegraphics[width=0.58\columnwidth]{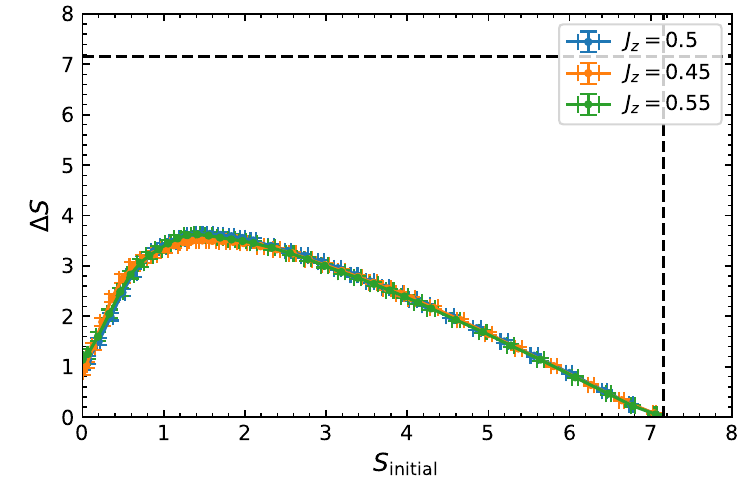}
    \caption{The HCEE growth as a function of the initial HCEE $S_{\text{initial}}$ for Hamiltonian MBL dynamics according to the setup described in the main text under different $J_z$ values. All data points are for system size $L=16$. The evolution time of saturated entanglement entropy is $t=10^{12}$. The data are averaged over 72 independent runs. Dashed lines mark the Haar measure average HCEE in the half-filling sector.}
    \label{prl_reply_different_Jz_MBL_Delta_HCEE}
\end{figure}

We have also investigated the influence of the anisotropy parameter $J_z$ on the entanglement dynamics. We performed new simulations with anisotropy values in $\hat{H}$ set to $J_z=0.45$ and $J_z=0.55$, and compared them to our original results at $J_z=0.5$. We have verified via level spacing ratio analysis that for these two new $J_z$ values, $\hat{H}$ still remains in thermal phase at $W=0.5$ and MBL phase at $W=5.0$, respectively. As presented in Fig.~\ref{prl_reply_different_Jz_MBL_Delta_HCEE}, the $\Delta S$ vs. $S_{\text{initial}}$ curves for all three values of $J_z$ collapse nearly onto a single, universal curve. This remarkable data collapse provides compelling evidence that the non-monotonic entanglement growth is a robust phenomenon, independent of the precise anisotropy.

\subsection{Universality Test with a Quantum Sun Model}

To test the universality of our findings beyond the disordered XXZ chain, we have adapted the ``quantum sun'' model from Ref.~\cite{PhysRevLett.129.060602}, a paradigm with established MBL phase. While the original model has a zero-dimensional geometry, we have mapped it to a one-dimensional chain to allow for a direct test within our framework, carefully preserving its key physical ingredients: a chaotic central quantum dot coupled with exponentially decaying interactions to a localized bath.

\begin{figure}[!htbp]
    \centering
    \includegraphics[width=0.5\columnwidth]{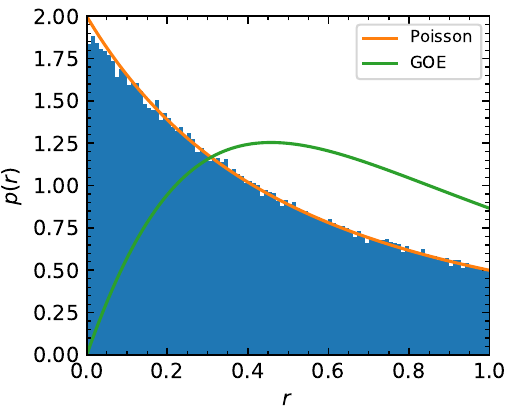}
    \caption{Level spacing ratio distributions $p(r)$ of $\hat{H}_\text{sun}$. For comparison, the theoretical predictions are overlaid as solid lines. The histogram is compiled from $500$ eigenstates in the middle of the energy spectrum of the entire Hilbert space of $500$ independent Hamiltonian realizations. 
    $L=14$. $\bar{r}\approx0.389$.}
    \label{prl_reply_quantum_sun_level_spacing}
\end{figure}

The Hamiltonian for our 1D adaptation, $\hat{H}_{\text{sun}}$, is given by:
\begin{equation}
\begin{aligned}
\hat{H}_\text{sun}=&\left(\prod_{i=1}^{\lfloor L_L/2\rfloor}\otimes\hat{I}\right)\otimes\hat{R}\otimes\left(\prod_{i=\lfloor L_L/2\rfloor+L_N+1}^{L}\otimes\hat{I}\right)\\
&+\sum_{i=1}^{\lfloor L_L/2\rfloor}\left(g_0\alpha^{u_i} \hat{S}_{n_i}^x\hat{S}_i^x+h'_i\hat{S}_i^z\right)+\sum_{i=\lfloor L_L/2\rfloor+L_N+1}^{L}\left(g_0\alpha^{u_i} \hat{S}_{n_i}^x\hat{S}_i^x+h'_i\hat{S}_i^z\right)
\end{aligned}
\label{eq:H_sun}
\end{equation}
where the quantum dot which comprises $L_N$ spins is located at the center of the chain, occupying sites indexed from $i=\lfloor L_L/2\rfloor+1$ to $i=\lfloor L_L/2\rfloor+L_N$, where $L_L$ is the number of spins outside the dot and $L_N+L_L=L$. $\hat{R}$ is an operator acting on the Hilbert space of the quantum dot and is drawn from the Gaussian orthogonal ensemble (GOE). $\hat{I}$ is the identity operator on the Hilbert space of a single spin. $u_i$ represents the distance between the quantum dot and the $i$-th spin outside the dot, it is determined by the following formula:
\begin{equation}
u_i=\begin{cases}2\left(\lfloor L_L/2\rfloor+1-i\right)+\zeta_i,&i\leqslant\lfloor L_L/2\rfloor\\2\left(i-\lfloor L_L/2\rfloor-L_N\right)-1+\zeta_i,&i\geqslant\lfloor L_L/2\rfloor+L_N+1\end{cases}
\label{eq:u_i}
\end{equation}
where $\zeta_i$ are drawn independently and uniformly from the interval $[-\zeta,\zeta]$ and $\zeta=0.2$. In this way, the relative magnitude of $u_i$ can be reflected in the tensor product order: for the spin on the left side of the quantum dot, $u_i$ decreases as $i$ increases; For the spin on the right side of the quantum dot, $u_i$ increases as $i$ increases. Therefore, the way we give a one-dimensional structure to the models in the literature is reasonable. Site $n_i$ is randomly selected within the dot. $h'_i$ are drawn independently and uniformly from the interval $[0.5,1.5]$. We set the constant parameters $g_0$ and $\alpha$ to be $1$ and $0.5$ respectively (see Fig.~\ref{prl_reply_quantum_sun_level_spacing} for level spacing ratio distribution).

\begin{figure}[!htbp]
    \centering
    \includegraphics[width=0.58\columnwidth]{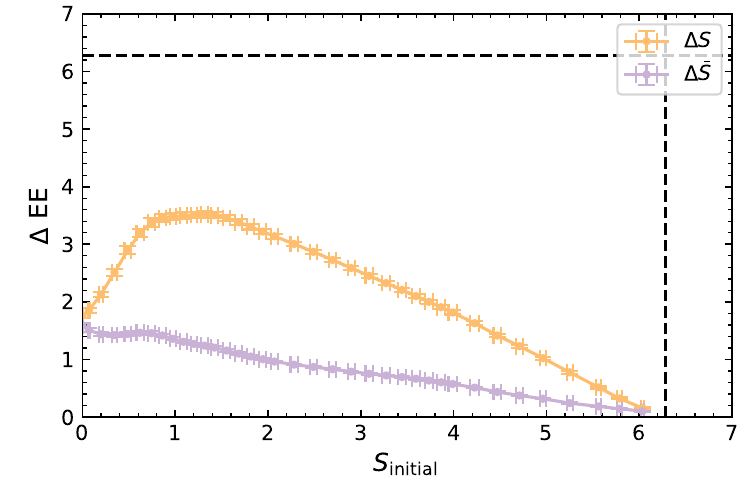}
    \caption{The HCEE growth and BAEE growth as a function of the initial HCEE $S_{\text{initial}}$ under the dynamics governed by $\hat{H}_\text{sun}$. All data points are for system size $L=14$. The evolution time of saturated entanglement entropy is $t=10^{12}$. The data are averaged over 72 independent runs. Dashed lines mark the Haar measure average HCEE in the whole Hilbert space, namely the Page value.}
    \label{prl_reply_quantum_sun_Delta_HCEE_Delta_BAEE}
\end{figure}

We then simulated the evolution of the HCEE growth ($\Delta S$) and BAEE growth ($\Delta \bar{S}$) starting from our partially thermalized states under this quantum sun-inspired model. The results, presented in Fig.~\ref{prl_reply_quantum_sun_Delta_HCEE_Delta_BAEE}, are qualitatively identical to those from our original disordered XXZ model. We again observe the characteristic non-monotonic peak in $\Delta S$, while the net entanglement generation, quantified by $\Delta \bar{S}$, remains small and decays almost completely monotonically. This agreement provides powerful evidence that the ``move-dominated'' nature of entanglement dynamics is a robust and universal feature of the MBL phase, not specific to a particular model.

\section{Other Build and Move Hybrid Dynamics Models}

In the main text, we classify thermalizing systems, free fermions, and certain classes of RQCs as ``build and move hybrid'' dynamics, all characterized by a monotonic decay of the HCEE growth with initial entanglement. To further test the breadth of this classification, we investigate two other paradigmatic models of quantum dynamics: the integrable XXZ chain and a chaotic dual-unitary circuit.

\subsection{Integrable XXZ Chain}

The connection between MBL and emergent integrability makes the study of conventional integrable models an important point of comparison. While our main text already includes the non-interacting free fermion model, we here present new results for the interacting integrable XXZ chain, which is obtained by setting the disorder strength $W=0$ in our Hamiltonian $\hat{H}$ (Eq.~\eqref{eqS1}).

\begin{figure}[!htbp]
    \centering
    \includegraphics[width=0.58\columnwidth]{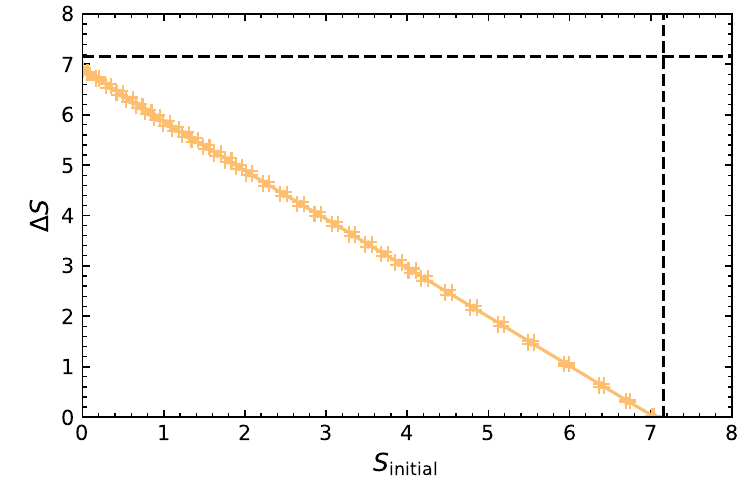}
    \caption{The HCEE growth as a function of the initial HCEE $S_{\text{initial}}$ under the dynamics governed by the integrable XXZ Hamiltonian. All data points are for system size $L=16$. The saturated entanglement entropy are determined by averaging over 10 late-time snapshots (evolution time $t=820.0, 840.0, 860.0, \cdots, 1000.0$). The data are averaged over 72 independent runs. Dashed lines mark the Haar measure average HCEE in the half-filling sector.}
    \label{prl_reply_XXZ_Delta_HCEE}
\end{figure}

The results for the HCEE growth starting from our partially thermalized states are shown in Fig.~\ref{prl_reply_XXZ_Delta_HCEE}. The integrable XXZ chain exhibits a monotonically decreasing $\Delta S$ as a function of $S_{\text{initial}}$, a behavior qualitatively similar to that of the free fermion and thermal cases. This confirms that integrable systems, both interacting and non-interacting, are classified as ``build-and-move hybrid'' within our framework. Their dynamics are fundamentally different from the non-monotonic, ``move-dominated'' behavior of MBL systems.

\subsection{Chaotic Dual-Unitary Circuit}

Dual-unitary circuits~\cite{PhysRevLett.123.210601,
PhysRevLett.125.070501,
10.21468/SciPostPhys.8.4.067,
PhysRevB.101.094304} are paradigms of maximal quantum chaos and serve as another important benchmark. For our RQC protocols, in the computational basis, the two-qubit unitary gate is given by
\begin{equation}
\begin{aligned}
U=\left[\begin{array}{cccc}
\mathrm{e}^{-\mathrm{i}\beta/4} & 0 & 0 & 0 \\
0 & \mathrm{e}^{\mathrm{i}\beta/4}\cos\frac\alpha2 & -\mathrm{i}\mathrm{e}^{\mathrm{i}\beta/4}\sin\frac\alpha2 & 0 \\
0 & -\mathrm{i}\mathrm{e}^{\mathrm{i}\beta/4}\sin\frac\alpha2 & \mathrm{e}^{\mathrm{i}\beta/4}\cos\frac\alpha2 & 0 \\
0 & 0 & 0 & \mathrm{e}^{-\mathrm{i}\beta/4}
\end{array}\right].
\end{aligned}
\label{eq:U}
\end{equation}
Its dual gate is given by
\begin{equation}
\begin{aligned}
\tilde{U}=\left[\begin{array}{cccc}
\mathrm{e}^{-\mathrm{i}\beta/4} & 0 & 0 & \mathrm{e}^{\mathrm{i}\beta/4}\cos\frac\alpha2 \\
0 & 0 & -\mathrm{i}\mathrm{e}^{\mathrm{i}\beta/4}\sin\frac\alpha2 & 0 \\
0 & -\mathrm{i}\mathrm{e}^{\mathrm{i}\beta/4}\sin\frac\alpha2 & 0 & 0 \\
\mathrm{e}^{\mathrm{i}\beta/4}\cos\frac\alpha2 & 0 & 0 & \mathrm{e}^{-\mathrm{i}\beta/4}
\end{array}\right].
\end{aligned}
\label{eq:dual_U}
\end{equation}
For $\tilde{U}$ to be unitary, we find that the condition is $\cos(\alpha/2) = 0$, which implies $\alpha=\pi$ for the parameter range we studied. This reveals that the gates used in our RQC(C), RQC(D), and random SWAP protocols are, in fact, dual-unitary. However, since they are located at the special trivial parameter points of the dual unitary quantum gate, they do not lead to thermalization.

\begin{figure}[!htbp]
    \centering
    \includegraphics[width=0.58\columnwidth]{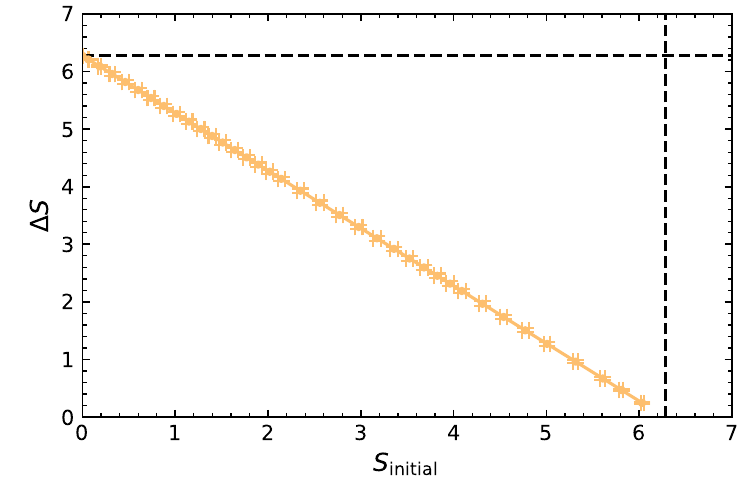}
    \caption{The HCEE growth as a function of the initial HCEE $S_{\text{initial}}$ under the random brick-layer circuit constructed of $U_{\text{SDKI}}$ and SWAP gates. All data points are for system size $L=14$. The evolution cycle number of saturated entanglement entropy is $2000$. The data are averaged over 72 independent runs. Dashed lines mark the Haar measure average HCEE in the whole Hilbert space, namely the Page value.}
    \label{prl_reply_DU_RQC_Delta_HCEE}
\end{figure}

To test a generic chaotic dual-unitary model, we performed new simulations using a dual-unitary gate from  Ref.~\cite{PhysRevLett.123.210601}, given by:
\begin{equation}
\begin{aligned}
U_{\text{SDKI}}=\frac{\mathrm{i}}{2}\left[\begin{array}{cccc}
-\mathrm{e}^{-2\mathrm{i}h} & \mathrm{e}^{-2\mathrm{i}h} & 1 & 1 \\
\mathrm{e}^{-2\mathrm{i}h} & \mathrm{e}^{-2\mathrm{i}h} & -1 & 1 \\
1 & -1 & \mathrm{e}^{2\mathrm{i}h} & \mathrm{e}^{2\mathrm{i}h} \\
1 & 1 & \mathrm{e}^{2\mathrm{i}h} & -\mathrm{e}^{2\mathrm{i}h}
\end{array}\right],
\end{aligned}
\label{eq:U_SDKI}
\end{equation}
which is dual-unitary for any real $h$ and is closely related to the self-dual kicked Ising model. We constructed a random brick-layer circuit with periodic boundary conditions where in each cycle, one randomly chosen bond is acted upon by $U_{\text{SDKI}}$ while the other $L-1$ gates in this cycle are SWAP gates. The results starting from our partially thermalized states show that this circuit leads to complete thermalization. As shown in Fig.~\ref{prl_reply_DU_RQC_Delta_HCEE}, the plot of HCEE growth $\Delta S$ exhibits a monotonic decay, qualitatively identical to the thermal RQC in our manuscript. This confirms that generic, chaotic dual-unitary circuits are significant examples of the ``build and move hybrid'' class, characterized by their efficient generation and transport of entanglement.

\bibliographystyle{apsreve} 
\bibliography{ref}